# High-performance cVEP-BCI under minimal calibration

Yining Miao[1], Nanlin Shi[1], Changxing Huang[1], Yonghao Song[1], Xiaogang Chen[2], Yijun Wang[3], Xiaorong Gao[1,*]

1,*Department of Biomedical Engineering, School of Medicine, Tsinghua University, Beijing, 100084, China.

2, Institute of Biomedical Engineering, Chinese Academy of Medical Sciences and Peking Union Medical College, Street, Tianjin, 300192, China.

3, State Key Laboratory on Integrated Optoelectronics, Institute of Semiconductors, Chinese Academy of Sciences, Beijing, 100083, China.

*email: gxr-dea@mail.tsinghua.edu.cn

## Abstract

The ultimate goal of brain-computer interfaces (BCIs) based on visual modulation paradigms is to achieve high-speed performance without the burden of extensive calibration. Code-modulated visual evoked potential-based BCIs (cVEP-BCIs) modulated by broadband white noise (WN) offer various advantages, including increased communication speed, expanded encoding target capabilities, and enhanced coding flexibility. However, the complexity of the spatial-temporal patterns under broadband stimuli necessitates extensive calibration for effective target identification in cVEP-BCIs. Consequently, the information transfer rate (ITR) of cVEP-BCI under limited calibration usually stays around 100 bits per minute (bpm), significantly lagging behind state-of-the-art steady-state visual evoked potential-based BCIs (SSVEP-BCIs), which achieve rates above 200 bpm. To enhance the performance of cVEP-BCIs with minimal calibration, we devised an efficient calibration stage involving a brief single-target flickering, lasting less than a minute, to extract generalizable spatial-temporal patterns. Leveraging the calibration data, we developed two complementary methods to construct cVEP temporal patterns: the linear modeling method based on the stimulus sequence and the transfer learning techniques using cross-subject data. As a result, we achieved the highest ITR of 250 bpm under a minute of calibration, which has been shown to be comparable to the state-of-the-art SSVEP paradigms. In summary, our work significantly improved the cVEP performance under few-shot learning, which is expected to expand the practicality and usability of cVEP-BCIs.

## Introduction

A brain-computer interface (BCI) enables users to communicate with the outside world by measuring and analyzing the brain activities (Gao et al., 2014). In comparison to the invasive version, BCIs based on visual evoked potential (VEP) promise to provide a high-speed, non-invasive communication experience based on electroencephalography (EEG) (Chen et al., 2015b; Wang et al., 2008). The ultimate goal for non-invasive BCIs at this stage is to achieve high performance under few or zero-shot learning, thus making them beneficial for daily plug-and-play communication among disabled individuals (Peters et al., 2020). While extensive research efforts have predominantly focused on relying on steady-state visual evoked potential BCIs (SSVEP-BCIs) due to their remarkable communication speed (Liu et al., 2021; Nakanishi et al., 2018) and, notably, their training-free characteristics (Bin et al., 2009c; Chen et al., 2015a, 2021), a recent study has highlighted the inefficiency of SSVEP-BCIs in utilizing the spectral resources of the primary visual pathway (Shi et al., 2023b.) This limitation leads to bottlenecks in communication speed and the number of encoding targets. In contrast, code-modulated VEP (cVEP)-based BCIs based on broadband white noise (WN) have been proposed to overcome the restrictions of SSVEP-BCIs and hold the promise of being the next-generation visual BCI system (Bin et al., 2009a; Shi et al., 2023b). However, due to the significant complexity of the spatial-temporal dynamics of broadband stimulation, achieving high performance in cVEP-BCIs typically comes with an overwhelming calibration effort (Bin et al., 2009b; Martínez-Cagigal et al., 2021). Therefore, minimizing the calibration effort of cVEP-BCIs has become an essential challenge in visual BCI research, with the potential to significantly enhance the practicality and user experience of BCI systems.

The necessity for calibration in cVEP-BCIs arises because our understanding of spatial-temporal dynamics is insufficient without calibration data. Unlike the intuitive frequency-following characteristics of SSVEP, the stimulus-response function under broadband stimulation becomes significantly more complex. As a result, it typically requires subject-dependent calibration data for learning to extract the spatial-temporal patterns. In SSVEP target identification problems, the temporal patterns can be approximated either by temporal patterns extracted from subject-dependent calibration data, or by sine and cosine waves at the stimulation frequencies. Therefore, both calibration-dependent and calibration-free decoding methods have been extensively studied for SSVEP-BCIs, with the highest information transfer rate (ITR) reaching 251.8 bpm using the task-discriminant component analysis (TDCA) proposed by Liu et al. in a calibration-based setting (Liu et al., 2021), and 151.18 bpm using the filter bank canonical correlation analysis (FBCCA) proposed by Chen et al. in a calibration-free setting (Chen et al., 2015). Up till now, the SSVEP-BCIs remain the only zero- or few-shot option that maintains high-speed performance with an ITR above 150 bpm among all alternatives. Conversely, although multiple researchers have worked on the zero/few-shot cVEP-BCI systems, most of their performance is still not comparable with SSVEP-BCIs. For example, Nagel et al. proposed the Code2EEG model in 2018, a linear regression method to estimate the temporal pattern of broadband stimulation from stimulus sequence. This approach reached an ITR of 108 bpm under a calibration effort of 384 s . Furthermore, Thielen et al. employed a similar approach to estimate the temporal pattern under short and long flashes. This approach achieved the ITR at 74 bpm and held the promise of working under zero calibration (Thielen et al., 2021). The underwhelming performance of cVEP-BCIs compared to SSVEP-BCIs is primarily attributed to the inefficiency and inaccuracy of broadband spatial-temporal pattern acquisition with insufficient calibration data.

Other than acquiring spatial-temporal patterns by modeling the stimulus-encoding process, the patterns can also be constructed through cross-subject data based on transfer learning techniques, which, unfortunately, have seldom been applied to cVEP studies. For SSVEP-BCIs, several studies have proposed transfer learning algorithms to reduce calibration efforts and enhance generalizability. For instance, Yuan et al. introduced the transfer template-based canonical correlation analysis (tt-CCA) method, acquiring temporal patterns by averaging cross-subject EEG data, resulting in a noteworthy 32.05% enhancement in target identification accuracy at 1.0 s compared to the standard CCA method (Yuan et al., 2015). Additionally, Wong et al. proposed the subject transfer-based CCA (st-CCA) method, which involves weighted averaging across subjects to acquire temporal patterns, thereby further enhancing the ITR (Wong et al., 2020). Shi et al. proposed a representative-based cold start (RCS) method, achieving an ITR of over 200 bpm by using transferable cross-subject data (Shi et al., 2023a). In theory, though not yet studied, it's reasonable to expect that broadband cVEP responses can also be transferable across individuals. Consequently, transfer learning methods are poised to play a significant role in enhancing the precision of spatial-temporal pattern construction in cVEP decoding.

The aim of this study is to develop a high-speed c-VEP BCI under minimal calibration. First, we introduce a single-target calibration phase lasting one minute to efficiently extract the spatial-temporal pattern. Subsequently, based on calibration data, we first employ the linear modeling method to construct temporal patterns from stimulus, which can be generalized to accommodate various stimulus patterns. We further incorporate subject-independent knowledge to construct the temporal pattern based on cross-subject data. Through a series of offline and online experiments, we demonstrated that the transfer learning method yields the highest average ITR of 250.16 bpm with only one-minute calibration. And most importantly, we validated that the cVEP-BCIs can reach comparable performance with the state-of-the-art SSVEP-BCIs under a few-shot learning scenario, with the former reaching a highest ITR of 177.90±51.77 bpm and the latter 182.63±58.28 bpm in the offline dataset. Considering the advantages of cVEP-BCIs, we discuss their future development in large-scale target identification as well as asynchronous BCI systems. In summary, this study develops a high-performance cVEP-BCI with minimal calibration effort, potentially facilitating BCI usage among a broader demographic.

## Methods and Materials

### 1. Stimulus Design

We implemented the newly-proposed WN stimuli as the cVEP modulation sequences (Shi et al., 2023b). For the calibration stage, we generated 20 classes of sequences encoded with uniformly distributed random white noise ranging from 0 to 1. The contrast level was then scaled and projected within the range of 0 to 255 for display purposes. For the target identification stage, we selected 40 classes of WN stimuli from an entire code space containing 10,000 WN sequences, which are all different from those used for calibration. The code selection was carried out using the simulating annealing algorithm (Ye et al., 2022), which aims to maximize the minimum pairwise Euclidean distances among codes. Importantly, the selection was conducted at the group level, where each individual adopted the identical stimulation paradigm. Additionally, we performed layout optimization to enhance the distinctiveness between neighboring targets, achieving maximum separability (Thielen et al., 2015).

To compare the few-shot performance with the SSVEP paradigm, we also included the state-of-the-art joint frequency phase modulation (JFPM) design (Chen et al., 2015b) in the target identification stage. The JFPM paradigm involves sampling sinusoidal stimulation across frequencies ranging from 8 Hz to 15.8 Hz with a 0.2 Hz interval and the corresponding phases starting from 0 and progressing in intervals of 0.5 $\pi$.

During the calibration step, the stimuli were presented in turn on a single target with a size of 200 × 200 pixels. A small red dot was displayed at the center of the flicker to help subjects focus their attention. During the target identification step, 40 visual flickers were arranged in a 5 × 8 matrix on the screen. Each visual flicker had a size of 140 × 140 pixels, and the horizontal and vertical distances between neighboring flickers were kept at 50 pixels. All flickers were displayed on a 1920 × 1080 LCD monitor with a 60-Hz refresh rate.

## 2. Experiment Protocol

### 2.1. Offline Experiment

The offline BCI experiment aims to validate the proposed methods on the few-shot cVEP-BCIs and make comparisons with SSVEP paradigms. 15 subjects (8 females) participated in this phase during the whole experiment. All subjects were required to read and sign an informed consent form approved by the Research Ethics Committee of Tsinghua University.

The experiment consisted of two stages: the calibration and target identification stages. During the calibration stage, participants are required to fixate at the center of a single target flicker modulated by 20 classes of WN sequences different from the target identification stage; each class is repeated for 4 trials, and each trial lasts for 3s with a 1s interval. Compared to the conventional calibration phase, where participants are required to move their gaze as the system instructs, the single-target calibration proposed in this study gains two obvious advantages: the first is that the system is calibrated without the need to alter the gaze on the multi-target speller; the second is that for the first time, the calibration stimulus and the stimulus used for identification are different, which means we can capture the invariant representation of the primary visual system and can generalize to multiple stimulus sequences and paradigms. This design was able to improve the simplicity of calibration and simultaneously facilitate the acquisition of robust spatial-temporal patterns.

Following the single-target calibration stage, each subject was instructed to perform both SSVEP and cVEP speller experiments. In the target identification stage, each of the 40 targets flickered for 3 s with a 0.5-s interval, and this pattern repeated across 5 blocks. Subjects were instructed to shift their visual attention to the cued target before stimulation started and to avoid eye blinks during stimulation. To balance the performance drift caused by fatigue, the order of the two alternatives was randomized.

The dataset obtained from the offline experiment was also used for the online experiment.

The EEG data were recorded using a Synamps2 system (NeuroScan, Inc.) at a 1000-Hz sampling rate with 62 channels in the offline experiment, while 21 of them (Pz, P1/2, P3/4, P5/6, P7/8, POz, PO3/4, PO5/6, PO7/8, Oz, O1/2, and CB1/2), optimized from the offline analyses, were used in the online experiment. The experiment was implemented in a shielded room with the distance between subjects and the screen being 65 cm and the contact impedance maintained below 15 $k\Omega$. After acquisition, the data were down-sampled to 250 Hz and notch-filtered at 50 Hz using an IIR filter.

## 2.2. Online experiment

The online experiment aims to evaluate subjects' online spelling performance of cVEP-BCI with real-time feedback. 10 subjects (8 females) were recruited, of which 7 overlapped with the offline experiment. For the recalled subjects, their data from offline experiments was excluded from the cross-subject data for the transfer learning method. Likewise, the online experiment also included the calibration and target identification stages. The procedures of the calibration step were the same as those of the offline experiment, except that the single target was modulated by a minimum of 5 classes of WN sequences, each repeated 4 trials with a stimulation time of 3 s, leading to a total calibration time of 1 minute. The target identification step consisted of both cued-spelling and free-spelling phases, employing linear modeling and transfer learning methods, respectively. Five best subjects selected from the cued-spelling phase were instructed to further participate in the free-spelling phase. During the experiments, the stimulation lasted 2 s for the linear modeling method and 0.75 s for the transfer learning method. Each method contained five blocks. In the free-spelling stage, subjects were required to type the phrase 'few shots wn bci' five times without visual cues for each method. The online experiments were conducted on separate days following the completion of the offline experiments. All parameters used in the online experiment were optimized based on the results obtained from offline analyses.

## 3. Methods for zero/few-shot learning

Based on the calibration data collected from the single target phase, we introduce two methods: linear modeling and transfer learning to construct spatial-temporal patterns. In this study, we define the "zero/few-shot" scenario as no or little subject-dependent calibration data being used. In this context, the spatial-temporal pattern in the linear modeling method can be derived either from subject-independent averaged data (zero-shot) or from subject-dependent data (few-shot). However, the transfer method relies on both subject-dependent and subject-independent data, and therefore, it can only be applied to the "few-shot" scenario.

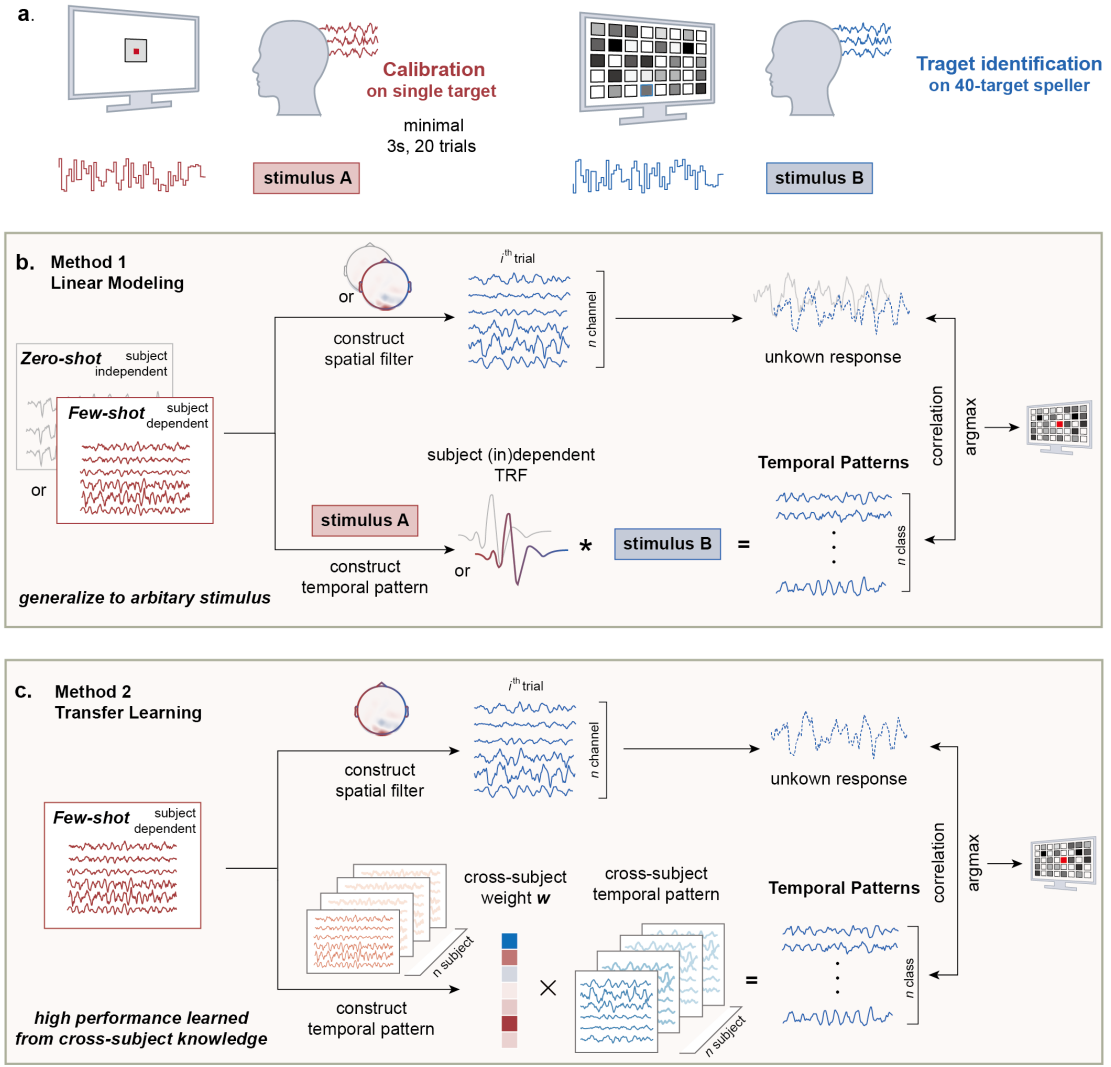

**Fig. 1 | Schematic view of the proposed methods. a.** The calibration phase (red) consist the single target stimulation modulated by WN sequence (denoted as stimulus A), then tested(blue) on the 40-speller modulated by a different set of WN sequence (denoted as stimulus B), **b. The linear modeling method,** where the spatial filter is acquired by either subject-dependent (red) or independent (gray) data, the temporal patterns are acquired from convolution between the stimulus and the TRF, **c, The transfer learning method,** where the spatial filter is acquired from subject-dependent data, and the temporal pattern is acquired by weighted cross-subject temporal patterns (denoted as light blue), the cross-subject weights (***w***) are learned from cross-subject calibration data (denoted as light red).

### 3.1. Spatial pattern based on TDCA

To extract the activation at the source level from the sensor-level data, we employed the TDCA method as the spatial filtering technique (Liu et al., 2021). The spatial filter, which can be interpreted as being related to the spatial activation during a specific task, should be generalizable across various stimulation sequences and layouts. Therefore, in this study, our objective is to utilize the spatial filters calculated during the calibration phase as substitutes for the multi-target spelling phase. This procedure is expected to significantly reduce the need for calibration.

Unlike the classical CCA (de Cheveigné et al., 2018; Liu et al., 2021; Thielen et al., 2021), or TRCA method (Nakanishi et al., 2018), which generate class-specific spatial filters, the TDCA method aims to derive class-generic spatial filters for all calibration data. This is achieved by computing the between-class scatter matrix $S_b$ and the within-class scatter matrix $S_w$ using multi-dimensional EEG signals encompassing multiple blocks and classes:

$$S_b = H_b H_b^T \tag{1}$$

$$S_w = H_w H_w^T \tag{2}$$

where $H_b$, $H_w$ is defined as

$$H_b = \frac{1}{\sqrt{N_c}} [\overline{X}^1 - \overline{X}, \dots, \overline{X}^{N_c} - \overline{X}] \tag{3}$$

$$H_w = \frac{1}{\sqrt{N_t}} [X^{(1)} - \overline{X}^{(1)}, \dots, X^{(N_t)} - \overline{X}^{(N_t)}] \tag{4}$$

where $\overline{X}^{(i)}$ and $\overline{X}^j$ are the two-dimensional cluster centers of the EEG signals from the $i^{th}$ trial and the $j^{th}$ class, respectively. Then, the Fisher criterion is employed to compute linear projecting vectors, aiming to minimize the divergence within classes and maximize the divergence between classes:

$$\hat{u} = \underset{u}{argmax} \frac{u^T S_b u}{u^T S_w u} \tag{5}$$

The source EEG signals $X_s$ can be generated by spatially filtering original EEG signals $X$ with $\hat{u}$, and then employed for template matching:

$$X_s = \hat{u}^T X \tag{6}$$

The spatial pattern $p$ is calculated as:

$$p = \Sigma_x u \Sigma_{\hat{s}}^{-1} \tag{7}$$

where $\Sigma_x$ and $\Sigma_{\hat{s}}$ are the covariance of the original EEG and estimated source EEG signals, respectively.

## 3.2. Temporal pattern based on linear modeling

In the linear modeling method, the temporal patterns are the estimated responses constructed from stimulation in the target identification stage. To obtain these estimated temporal patterns, we adopted the previous studies to model the stimulus-response function as a linear temporal filtering process. These linear temporal filters, commonly referred to as the temporal response function (TRF) (Crosse et al., 2021, 2016), are obtained from the calibration data. Based on the subject-dependent or independent calibration data used, the linear modeling method can be devised for zero or few-shot scenario. Under this linear time-invariant (LTI) system hypothesis (Crosse et al., 2021; Thielen et al., 2015), the EEG responses $r(t)$ can be calculated through the convolution of the stimulation input $s(t)$ and the TRF $h(\tau)$:

$$r(t) = \sum_{\tau} h(\tau) s(t - \tau) + \varepsilon(t) \tag{8}$$

where $\varepsilon(t)$ represents the reconstruction noise of the system.

The temporal filter $h(\tau)$ captures the invariant representation of neural dynamics, thereby ensuring generalizability across different stimulation sequences. The value of $h(\tau)$ can be acquired through least square estimation (LSE):

$$\hat{\boldsymbol{h}} = (\boldsymbol{S}^T\boldsymbol{S})^{-1}\boldsymbol{S}^T\boldsymbol{r} \tag{9}$$

where $\boldsymbol{S}$ is the Hermitian matrix of stimulus sequences, where each column represents the stimulus sequences delayed from $\tau_{min}$ to $\tau_{max}$, and $\boldsymbol{r}$ is the vectorized form of EEG responses, $\tau_{min}$ and $\tau_{max}$ are set to 0 and 0.5 s in this study, respectively.

To achieve a universal temporal filter reflecting source activities, the source EEG signals are concatenated across all stimulus conditions. Subsequently, a singular value decomposition (SVD) method is employed to address the ill-conditioned issues associated with WN sequences:

$$\tilde{\boldsymbol{S}}^T\tilde{\boldsymbol{S}} = \boldsymbol{C}_{\tilde{S}\tilde{S}} = \boldsymbol{U}\boldsymbol{\Lambda}\boldsymbol{U}^T, \boldsymbol{\Lambda} = diag(\lambda_1, \dots, \lambda_Y) \tag{10}$$

$$m = \underset{m}{argmax}\left(\frac{\lambda_1 + \lambda_2 + \cdots + \lambda_m}{\lambda_1 + \lambda_2 + \cdots + \lambda_Y} < \alpha\right) \tag{11}$$

$$\hat{\boldsymbol{h}}^* = \left(\boldsymbol{C}_{\tilde{S}\tilde{S}}^*\right)^{-1}\tilde{\boldsymbol{S}}^T\tilde{\boldsymbol{r}} = \boldsymbol{U}diag\left(\frac{1}{\lambda_1},\frac{1}{\lambda_2},\dots,\frac{1}{\lambda_m},0,\dots,0\right)\boldsymbol{U}^T\tilde{\boldsymbol{S}}^T\tilde{\boldsymbol{r}} \tag{12}$$

Eq. (10) ~ (12) illustrates equations to compute the robust TRF $\boldsymbol{h}^*$, where $\tilde{\boldsymbol{r}} = [\boldsymbol{r}_1^T, \boldsymbol{r}_2^T, \dots, \boldsymbol{r}_{N_c}^T]^T$, $\tilde{\boldsymbol{S}} = [\boldsymbol{S}_1^T, \boldsymbol{S}_2^T, \dots, \boldsymbol{S}_{N_c}^T]^T$, and $N_c$ is the number of stimulation classes. $\alpha$ is set to 0.9 in this study.

Based on the spatial filter acquired through the calibration stage and the temporal pattern reconstructed using $h(\tau)$, the identification process involves comparing the correlation coefficient $\rho$ between the spatially filtered response and each temporal pattern. The ultimate predicted result is generated by selecting the class corresponding to the maximum $\rho$ value. Within the template matching framework, we integrate two conventional data augmentation procedures to boost classification performance: template shifting and filter banks. Consequently, the correlation coefficients are computed according to the following equation:

$$\tilde{\rho}_i^l = \rho(X, Y_i^l) = \sum_{n=1}^{N_{fb}} w_{fb}(n) \cdot \left(\rho_i^{n,l}\right)^2, l = 0, \pm 1, \ldots, \pm N_l, i = 1, 2, \ldots, N_c \quad (13)$$

where the weight for each sub-band is defined as $w_{fb}(n) = n^{-a} + b, n \in [1, N_{fb}]$, where $N_{fb}$ refers to the number of filter banks, $N_l$ stands for the number of shifted sampling points, and $N_c$ represents the number of classes. Targets are identified using the following equation:

$$Q = \underset{i}{argmax}\ \underset{l}{max}\ \tilde{\rho}_i^l \quad (14)$$

### 3.3. Temporal pattern based on transfer learning

Unlike the previous method, the temporal patterns in the transfer learning method is constructed through the linear combination of the cross-subject data (Wong et al., 2020). The estimated temporal pattern for the $k^{th}$ stimulus $\tilde{x}_k$ is calculated as follows:

$$\tilde{x}_k = \frac{1}{N_{sub}} \sum_{n=1}^{N_{sub}} w_n \cdot \bar{X}_k^{(n)} \cdot v^{(n)} \quad (15)$$

where $\bar{X}_k^{(n)} \in \mathbb{R}^{N_t \times N_{ch}}$ denotes the $k^{th}$ class of averaged EEG response for the $n^{th}$ cross-subject, where $N_t$ represents the time-sampling number, $N_{sub}$ represents the cross-subject number. The spatial filter $v^{(n)}$ is also calculated by TDCA for the $n^{th}$ subject. The cross-subject weight $w_n$ reflects the similarity or transferability between subject dependent and independent data and can be calculated on the calibration data

by minimizing the error of the subject dependent response and the weighted cross-subject response:

$$w = (A^T A)^{-1} A^T b \tag{16}$$

where $w = [w_1, w_2, \cdots, w_{N_{sub}}]^T \in \mathbb{R}^{N_{sub} \times 1}$, $A$ and $b$ represent the concatenated version of spatially filtered temporal patterns for subject independent and dependent calibration data, respectively:

$$b = [u^T \overline{X}_1, u^T \overline{X}_2, \cdots, u^T \overline{X}_{N_c}] \tag{17}$$

$$A = \begin{bmatrix} \overline{\mathcal{X}}_1^{(1)} \cdot v^{(1)} & \overline{\mathcal{X}}_1^{(2)} \cdot v^{(2)} & \cdots & \overline{\mathcal{X}}_1^{(N_{sub})} \cdot v^{(N_{sub})} \\ \overline{\mathcal{X}}_2^{(1)} \cdot v^{(1)} & \overline{\mathcal{X}}_2^{(2)} \cdot v^{(2)} & \cdots & \overline{\mathcal{X}}_2^{(N_{sub})} \cdot v^{(N_{sub})} \\ \vdots & \vdots & \ddots & \vdots \\ \overline{\mathcal{X}}_{N_c}^{(1)} \cdot v^{(1)} & \overline{\mathcal{X}}_{N_c}^{(2)} \cdot v^{(2)} & \cdots & \overline{\mathcal{X}}_{N_c}^{(N_{sub})} \cdot v^{(N_{sub})} \end{bmatrix} \tag{18}$$

The subject dependent $\overline{X}_k$ and independent calibration data $\overline{\mathcal{X}}_k^{(n)}$ of the $n^{th}$ subject and $k^{th}$ class of responses are spatially filtered by respective filters $v^{(n)}$ and $u$, and then concatenated across all stimulus classes. The final results are identified through the same template matching procedures as the previous method.

The cross-subject weight $w$ represents the transferability across subjects, which is generalizable between cVEP and SSVEP paradigms. Therefore, the transfer learning method also works for SSVEP-BCIs by obtaining the weight $w$ through calibration data on broadband stimulation, and testing on the narrowband SSVEP for identification.

## 4. Performance Evaluation

In this study, we use classification accuracy and ITR to evaluate the target identification performance of BCI systems. The equation of the ITR calculation is as follows:

$$\text{ITR} = \left( \log_2 M + P\log_2 P + (1-P)\log_2\left[\frac{1-P}{M-1}\right]\right) \times \left(\frac{60}{T}\right) \text{(bpm)} \quad (19)$$

where $M$ is the target number, $P$ is classification accuracy, and $T$ is identification time (including 0.5-s gaze shifting time). Other than evaluating information transfer from the identification perspective, we also compute the mutual information $I$ by signal-to-noise ratio (SNR) analysis within the information theory framework (Shi et al., 2023b). To calculate the SNR in the frequency domain, the averaged spatially filtered temporal patterns $\bar{X}$ are considered to be the signal component, and the residual between which and the spatially filtered single trial data are considered the noise component:

$$\boldsymbol{N}_i = \boldsymbol{X}_i - \frac{1}{N_{tr}}\sum_i \boldsymbol{X}_i = \boldsymbol{X}_i - \bar{\boldsymbol{X}} \quad (20)$$

where $\boldsymbol{X}_i$ and $\boldsymbol{N}_i$ respectively represent the data and noise component for the $i^{\text{th}}$ trial, and $N_{tr}$ is the trial number. Finally, the mutual information is obtained by the spectral domain of the signal and noise component:

$$I = \int_0^k \log_2(1 + \text{SNR}(f))\, df \quad (21)$$

and

$$\text{SNR}(f) = \frac{\bar{X}(f)\cdot \bar{X}^*(f)}{\sum_i \boldsymbol{N}_i(f)\cdot \boldsymbol{N}_i^{\,*}(f)} \quad (22)$$

where $*$ represents the complex conjugate.

## Results

**1. Spatial-temporal patterns**

The results in Fig. 2(a)-(e) first show that the spatial-temporal pattern derived from the single-target calibration phase can be effectively applied to the multi-target identification phase. The subject-averaged TRFs shown in Fig. 2(a) in both phases exhibit similar patterns, typically characterized by two negative peaks and one positive peak at ~100 ms. However, the TRFs calculated in the 40-target speller paradigm are prone to show much smaller amplitudes and more jitters compared with the calibration settings, which is largely attributed to the influences of neighboring targets. Likewise, fig. 2(b) reveals a notable similarity in the spatial pattern, predominantly distributed in the occipital lobe across both phases.

For the transfer learning method, Fig. 2(c) demonstrates that the cross-subject model achieves modest predictability within just 1 s, which confirms the transferability between individual subjects. Furthermore, we compare the subject-variability of TRFs to uncover the underlying mechanisms of transfer learning. Fig. 2(d) illustrates that individual TRFs share similarities in the waveform but differ in amplitude and peak latency. After linear combination, the cross-subject data can yield temporal patterns similar to those generated by individual data (Fig. 2(d)). In summary, our findings demonstrate consistent spatial-temporal dynamics of the human visual system across diverse stimulation scenarios. This consistency underscores the feasibility of constructing efficient spatial-temporal patterns for classification with minimal calibration efforts.

Building upon the stable temporal filter $h(\tau)$, the results further show that individual temporal patterns can be approximated through reconstruction from stimulation and a linear combination of cross-subject responses. Fig. 2(e) and (f) respectively, depict the temporal and spectral representations of response patterns using the linear modeling and transfer learning methods. It's evident that both methods efficiently approximate the true response patterns (Fig. 2(e)). However, when observed from the spectral domain (Fig. 2(f)), it becomes apparent that the transferred temporal pattern, derived from the actual EEG response, captures the nonlinear components under broadband stimulation, particularly at higher frequency ranges (over 30 Hz). The results also suggest that the nonlinear dynamics in the higher frequency range exhibit consistency across individuals. On the contrary, the linear modeling method solely captures the linear component and peaks in the alpha and beta frequency ranges. Therefore, the transfer learning method based on actual responses yields more efficient temporal patterns than those obtained through stimulus reconstruction. This enhanced efficiency is anticipated to significantly contribute to the improvement of classification performance.

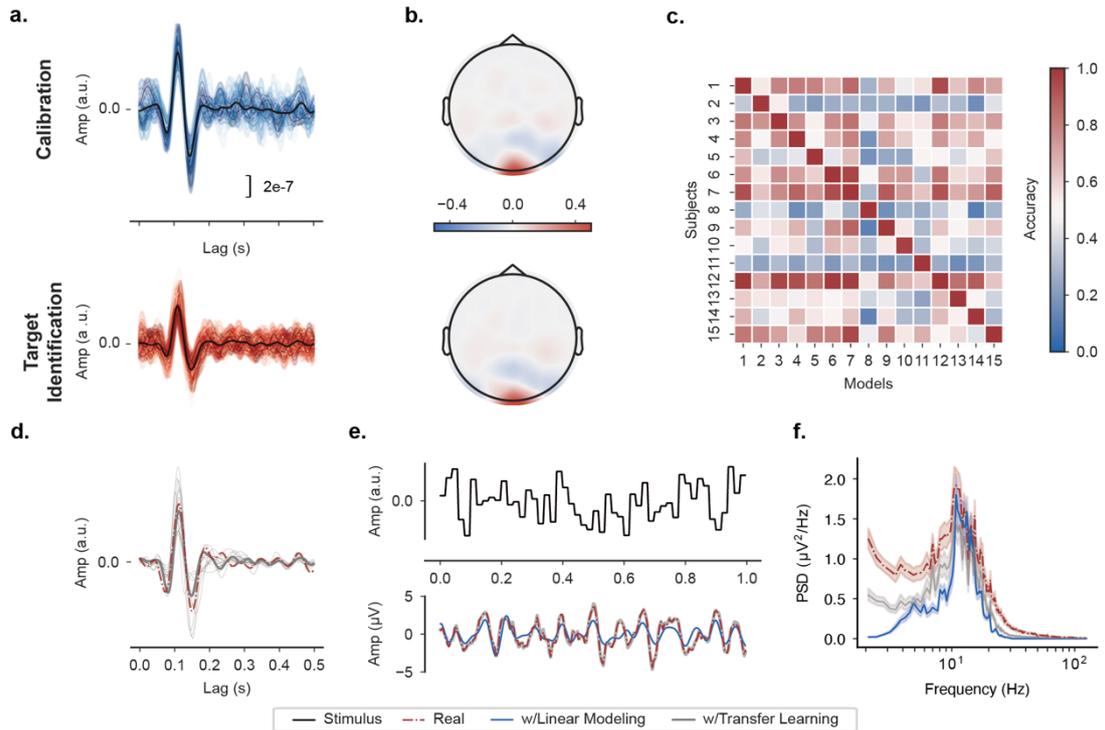

**Fig. 2 | Spatial-temporal patterns and cross-subject model predictability. a**, the subject-averaged TRFs from the single-target stage (blue lines) and the speller stage (red lines) using the linear modeling method (*n*=15, mean). Black lines represent the universal TRF across classes, others represent the TRF of each class. **b**, the subject-averaged spatial patterns from both stages (*n*=15). **c**, the cross-subject classification accuracy confusion matrix within 1.0 s. **d**, the individual and transferred TRFs. The red and gray lines depict the real and transferred TRF from a representative subject calculated using the subject-dependent and independent calibration data, respectively, the background lines depict TRFs from other source subjects. **e-f**, the subject-averaged temporal pattern and spectral representations of EEG responses (*n*=15, mean, 95% CI). The black line denotes a stimulus sequence, the red, blue, and gray lines denote the corresponding real response, the response constructed through linear modeling, and through transfer learning respectively.

## 2. Offline BCI Performance

The classification results demonstrate the high performance achieved by both methods.

In the linear modeling method, it can be seen from Fig. 3(a) that the highest average ITR of 90.71±23.19 bpm is attained at 2.0 s for WN cVEP under few-shot learning. Notably, even under zero-shot conditions where temporal patterns are formed using subject-averaged $h(\tau)$, the highest ITR reaches 65.35±31.68 bpm at 2.25 s, indicating the practicability of calibration-free WN cVEP-BCI (Fig. 3(b)). The incorporation of few-shot learning notably enhances ITR (P<0.001 by two-way ANOVA), underscoring the role of the single-target calibration phase in determining subject-specific spatial and temporal patterns. However, SSVEP-BCI based on FBCCA consistently outperforms WN cVEP-BCI employing the linear modeling method, achieving a significantly higher average ITR of 134.38±34.95 bpm at 1.25 s (Fig. 3(c), P<0.001 by two-way ANOVA). This performance gap is mainly attributed to FBCCA capturing the nonlinear harmonic components of SSVEP, whereas the linear modeling method only harnesses the linear components of WN cVEP, which are primarily concentrated within the 10-20 Hz bandwidth. The mechanism behind the performance gap between linear and nonlinear components can be further validated when compared with the typical CCA method, which only utilizes stimulation waveforms without harmonics as temporal templates. The performance of SSVEP-BCI with CCA (Fig. 3(d)) yields similar results with zero calibration WN cVEP based on subject-averaged $h(\tau)$ (Fig. 3(b), NS by two-way ANOVA). The comparison between these four conditions indicates that both individual knowledge and the nonlinear components are essential for constructing effective temporal templates, consequently affecting the identification performance.

In the transfer learning method, both paradigms yield comparable performances, with WN cVEP gaining a highest ITR of 177.90±51.77 bpm and SSVEP reaching 182.63±58.28 bpm at 0.75 s, respectively (NS by two-way ANOVA, Fig. 3(e) and (f)). However, when considering the top 5 subjects' performance, WN cVEP significantly outperforms SSVEP, with the highest average ITR of 255.78±39.22 bpm at 0.5 s and 196.56±51.77 bpm at 0.75 s, respectively (P=0.0014 by two-way ANOVA, subgraphs of Fig. 3(e) and (f)). These results highlight that the transfer learning method excels in constructing more efficient temporal patterns compared to the linear modeling method, leading to superior performance. Overall, we confirm that WN cVEP-BCI can achieve high speeds comparable to SSVEP-BCI under few-shot learning.

The comparative results of SNR and mutual information in Fig. 3(g) and (h) further highlight that the primary factor influencing the close BCI performance of WN cVEP and SSVEP is their similar utilization of band resources. The SNR curve in Fig. 3(g) offers a clear illustration of the spectral distinctions between WN cVEP and SSVEP. SSVEP exhibits an especially high SNR around 10 Hz owing to its resonance interaction with intrinsic alpha rhythms, while the high- SNR band is constrained between 8 Hz and 15.8 Hz due to the stimulus frequency band range limitation. In contrast, WN cVEP boosts a broader high SNR bandwidth attributed to its wider range of stimuli. Overall, SSVEP shows a higher SNR peak while WN cVEP demonstrates a more balanced SNR distribution, resulting in a close utilization of band resources and, consequently, similar mutual information (NS by t-test at 125 Hz, Fig. 3(h)). Since mutual information is closely related to the ITR (Shi et al., 2023b), these two paradigms exhibit similar BCI performance. Furthermore, the superior performance of WN cVEP over SSVEP for the top 5 subjects suggests that the WN cVEP-BCI system possesses a greater information transmission capacity than the SSVEP-BCI system for these individuals.

Based on the high BCI performance WN cVEP can reach, the comparative results of temporal filters and ITRs in Fig. 3(i)-(l) further demonstrate that the calibration effort required for WN cVEP-BCI is quite little. The subject-averaged temporal filters depicted in Fig. 3(i) and (j) show that the amplitude of both individual and cross-subject TRFs increases with longer calibration time, stabilizing at approximately 36 s. Regarding BCI performance, the results in Fig. 3(k) and (l) indicate significant improvements when the calibration time increases from 9 s to 27 s, but no obvious changes with continuous increments in training time for both methods ($P<0.001$ at 9 s, $P=0.016$ at 18 s, NS $\geq 27$ s for the linear modeling method, $P<0.001$ at 9 s, $P=0.05$ at 18 s, NS $\geq 27$ s for the transfer learning method, by t-test). Therefore, the minimum calibration effort required for a few-shot WN cVEP-BCI is 27 s.

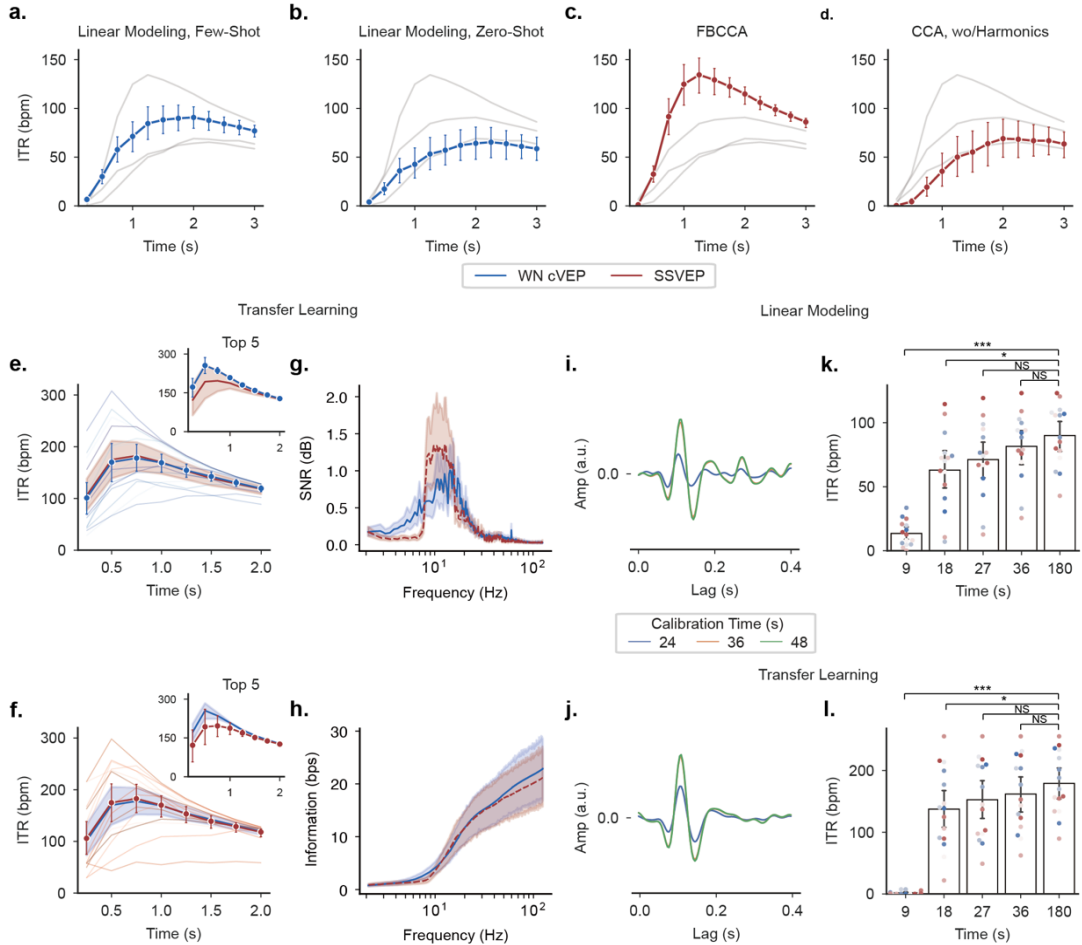

**Fig. 3 | Offline BCI Performance**. **a-b**, the subject-averaged ITRs of few/zero-shot WN cVEP-BCI using the linear modeling method ($n$=15, mean, 95% CI). **c-d**, the subject-averaged ITRs of SSVEP-BCI using (FB)CCA without (with) harmonics ($n$=15, mean, 95% CI). **e-f**, the ITRs of few-shot WN cVEP-BCI and SSVEP-BCI using the transfer learning method (mean, 95% CI, NS when $n$=15, P=0.0014 for the top 5 subjects by two-way ANOVA). **g-h**, the subject-averaged SNR and mutual information of temporal templates constructed using the transfer learning method ($n$=15, mean, 95% CI, NS by t-test when $f$=125 Hz for mutual information). **i-j**, the subject-averaged TRFs with increasing calibration time ($n$=15, above for the linear modeling method, below for the transfer learning method). **k-l**, the subject-averaged ITRs with increasing calibration time ($n$=15, mean, 95% CI, P<0.001 at 9 s, P=0.016 at 18 s, NS ≥27 s for the linear modeling method above, P<0.001 at 9 s, P=0.05 at 18 s, NS ≥27 s for the transfer learning method below, by t-test)

## 3. Online BCI Performance

Both the linear modeling and transfer learning methods are validated through an online BCI experiment, including cued-spelling and free-spelling tasks. The subjects involved in the online experiment overlap with those from the offline, and the data from the overlapping subjects is not transferred. According to the offline analyses, stimulation durations of 2.0 s and 0.75 s are respectively set for the linear modeling and transfer learning methods in the target identification stage. The calibration time is set to 60 s during the online experiments in order to get more stabilized spatial-temporal patterns. The optimized parameters are uniformly applied to all subjects in both methods. The results of cued-spelling BCI performance are presented in Table 1. The outcomes reveal an average accuracy of 87.8 ± 12.59% with the linear modeling method and 84.40 ± 16.80% with the transfer learning method, leading to an average ITR of 101.52±21.64 bpm and 193.17±58.39 bpm, respectively. Table 2 shows the results of free-spelling performance in 5 subjects. The findings demonstrate an average accuracy of 97.00±3.76% and 99.00±2.00% with the linear modeling and transfer learning methods, respectively, yielding an average ITR of 120.25±9.32 bpm and 250.16±10.57 bpm. The highest ITR of 127.73 bpm and 255.45 bpm are achieved with the two methods, respectively.

Table 1. Online Cued-Spelling Few-shot BCI Performance

|         | w/Linear modeling | | w/Transfer learning | |
|---------|---------|-----------|---------|-----------|
| Subject | Acc (%) | ITR (bpm) | Acc (%) | ITR (bpm) |
| S1 | 93.50  | 111.15 | 94.50  | 226.75 |
| S2 | 100.00 | 127.73 | 92.00  | 215.85 |
| S3 | 91.00  | 105.83 | 100.00 | 255.45 |
| S4 | 92.50  | 108.99 | 94.50  | 226.75 |
| S5 | 96.50  | 118.03 | 98.00  | 243.59 |
| S6 | 82.50  | 89.47  | 47.00  | 73.12  |
| S7 | 88.00  | 99.80  | 79.00  | 166.58 |
| S8 | 92.00  | 107.93 | 63.00  | 115.95 |

|        | 　     |        |        |        |
|--------|--------|--------|--------|--------|
| S9     | 52.50  | 43.52  | 77.50  | 161.45 |
| S10    | 89.50  | 102.78 | 98.50  | 246.25 |
| Mean±STD | 87.80±12.59 | 101.52±21.64 | 84.40±16.80 | 193.17±58.39 |

Table 2. Online Free-Spelling Few-shot BCI Performance

|         | Linear modeling | | Transfer learning | |
|---------|---------|---------|---------|---------|
| Subject | Acc (%) | ITR (bpm) | Acc (%) | ITR (bpm) |
| S1 | 100.00 | 127.73 | 95.00 | 229.02 |
| S2 | 100.00 | 127.73 | 100.00 | 255.45 |
| S3 | 91.25 | 106.35 | 100.00 | 255.45 |
| S4 | 93.75 | 111.70 | 100.00 | 255.45 |
| S5 | 100.00 | 127.73 | 100.00 | 255.45 |
| Mean±STD | 97.00±3.76 | 120.25±9.32 | 99.00±2.00 | 250.16±10.57 |

# Discussion

Focusing on the high-performance WN cVEP-BCI under minimal calibration, we first propose a single-target stage to extract individual spatial-temporal patterns within minute calibration. Building upon the temporal dynamics derived from stimulus sequences and cross-subject data, we introduce two decoding methods based on linear modeling and transfer learning, respectively. The results show that cVEP-BCIs can also achieve comparable performance to SSVEP-BCIs with minimal calibration. In addition, there are promising opportunities for further refinement and application of the proposed methods.

**1. Methods comparison**

The two methods introduced in this study offer distinct pros and cons, as outlined in Table 3. The transfer learning method outperforms the linear modeling method in terms of BCI performance owing to its greater efficacy in temporal pattern construction based on real responses. However, it's important to note that the transfer learning method comes with more limitations in BCI applications compared with the linear modeling one, since the former requires datasets collected under specific stimulus conditions. On the contrary, the linear modeling method does not necessitate any prior data, offering greater flexibility in system implementation as well as advantages for research on novel stimulus paradigms. Furthermore, the transfer learning method can still yield underwhelming results if the spatial-temporal patterns of cross-subject data have huge disparities, as witnessed in S6 in the online experiment. The comparative results between the proposed methods emphasize the significance of pattern construction efficacy in influencing target identification performance. Considering the substantial contributions that online data adaptation has shown in improving BCI performance (Wong et al., 2022), an approach worth exploring involves iteratively computing spatial-temporal patterns using individual actual responses from previous trials during the target identification stage. The adaptation process may further enhance the efficacy of pattern construction in real scenarios, ultimately resulting in even higher performance.

Table 3. Comparison between the Proposed Methods

| Methods/Items | Linear modeling | Transfer learning |
|---|---|---|
| Spatial pattern | Single-target calibration ||
| Temporal pattern | Stimuli reconstruction | Cross-subject responses |
| Performance | Low (120.25 bpm) | High (250.16 bpm) |
| Minimum calibration | 60 s ||
| Restriction | None | Cross-subject data under Specific Stimulus Condition |

## 2. Future Developments

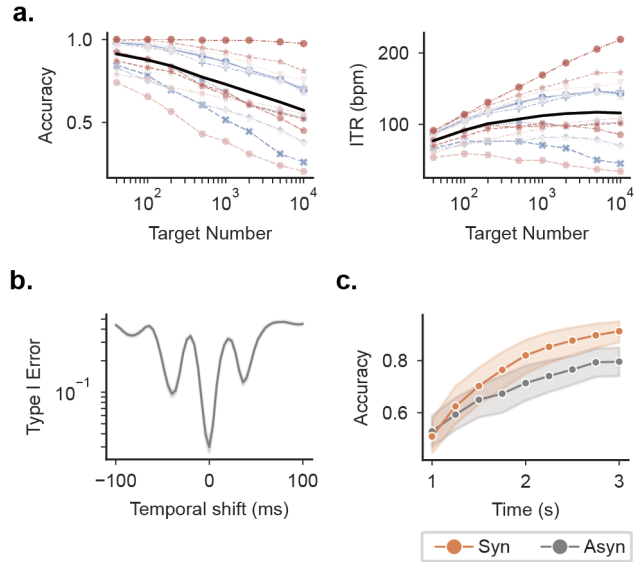

**Fig. 4 | Future development of few-shot WN cVEP-BCI. a**, the accuracy and ITR curves with the increment of the target number at 3.0 s (*n*=15). The background lines represent individual results, the black line represents the subject-averaged results. **b**, the subject-averaged type I error curve with the variance of temporal shift from -100 to 100 ms (*n*=15, mean, 95% CI). **c**, the subject-averaged accuracy curves of the synchronous (orange) and asynchronous (gray) system (*n*=15, mean, 95% CI).

First, the calibration burden can be further reduced by shifting broadband stimulus. In our study, we independently generated WN stimuli that were uncorrelated with each other. However, for cVEP-BCIs based on classical m-sequences, as proposed by Sun et al. in 2022, they encoded 120 targets using four 31-bit pseudorandom codes with cyclic shifts. This design significantly reduced the calibration time to less than 5 minutes and achieved an impressive ITR exceeding 250 bpm. Theoretically, future studies can explore the application of the cyclic shift technique to WN-based BCIs, further reducing the calibration efforts.

Second, the few-shot performance of cVEP BCIs can be further enhanced by increasing the target number. As many recent studies aim to increase the target number in order to boost ITR (Chen et al., 2022; Nagel and Spüler, 2019, 2018), WN cVEP shows clear flexibility in creating visual BCI spellers with a large number of targets. Combined with the linear modeling method, it is possible to perform calibration-free identification with infinite stimulation classes. To evaluate the performance with an increment of target number, we conducted a simulated experiment involving matching 40 classes of real EEG responses to $n$ ($n = 40, 100, 200, 500, 1000, 2000, 5000, 10000$) classes of templates constructed by WN sequences (Nagel and Spüler, 2018). The results in Fig. 5(a) implies that the subject-averaged ITR increases from $77.03\pm12.08$ bpm to $115.98\pm50.89$ bpm when the target number reaches 10000, and the best subject achieves an accuracy of 97.5% and an ITR of 219.20 bpm under the 10000-target classification. 4 out of 15 subjects still achieve increasing ITRs even with a very large target number (~10000), suggesting the efficacy of more-than-100-class identification and the potential for achieving even higher ITRs by increasing the target number. In fact, compared to SSVEP modulated with frequency division multiple access (FDMA) techniques, cVEP modulated with code division multiple access (CDMA) techniques is more suitable for encoding thousands or even millions of targets, as it utilizes spectrum resources more evenly and efficiently (Shi et al., 2023b).

Third, an asynchronous WN cVEP-BCI system is able to be implemented using the linear modeling method. In algorithms with calibration, synchronized event triggers are crucial for accurate spatial-temporal pattern extraction through phase-locking among different trials. However, by predicting response waveforms from stimulus onsets, we can identify the onset positions of stimuli in real EEG signals by matching the shifted templates. This allows for the creation of an asynchronous WN cVEP-BCI system independent of synchronized event triggers, improving system portability. To estimate the effects of trigger-free system, we calculate the average $1^{st}$ type error probability across all trials while varying template shifts (from -100 to 100 ms), as shown in Fig. 5(b). The Type I error is computed using a t-test to test the statistical significance of the largest correlation coefficient among the distribution of the other coefficients at each shifting point. It can be observed that the minimum probability value occurs at a shift of 0 ms, indicating the phase-locking point, which is consistent with the actual situation. Fig. 5(c) compares the ITRs of synchronous and asynchronous systems, demonstrating the feasibility of the trigger-free system with an average accuracy of 79.6% at 3.0 s, albeit with slightly lower performance compared with the synchronous system.

Despite the decent performance that the linear modeling method can achieve in zero/few-shot WN cVEP-BCI, the plateauing of performance with more than 1000 targets and the drop in accuracy with the asynchronous system highlight the need for a more comprehensive understanding of the dynamic characteristics of the visual pathway. To address this, nonlinear components should be taken into consideration to avoid information leakage during target identification.

## 3. Reconstruct SSVEP from $h(\tau)$

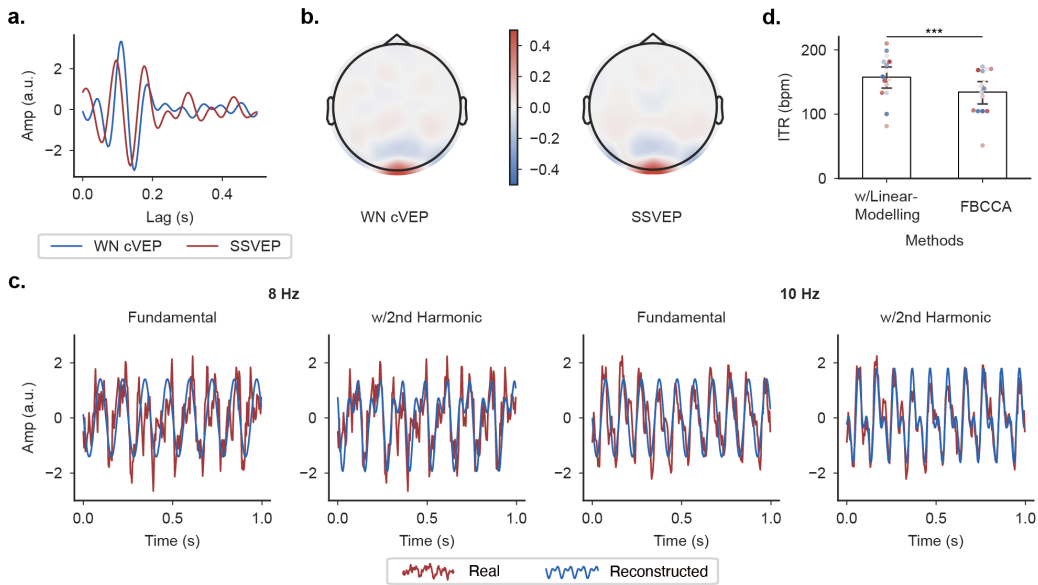

**Fig. 5. | SSVEP reconstruction**. **a**, the subject-averaged TRFs of WN cVEP and SSVEP paradigms (*n*=15). **b**, the subject-averaged spatial patterns of both paradigms (*n*=15). **c**, the waveforms of 8 Hz and 10 Hz real SSVEP and reconstructed SSVEP with fundamental and 2$^{nd}$ harmonic frequencies from a representative subject. **d**, the ITRs of SSVEP-BCI using the linear modeling method and FBCCA (*n*=15, mean, 95% CI, P<0.001by two-way ANOVA).

Several studies have proposed to decrease the calibration effort of SSVEP-BCI by constructing temporal patterns from sine/cosine stimuli (Wang et al., 2023; Wong et al., 2021). The broadband single-target flicker used in this study provides a much more efficient way to calibrate, even for SSVEP BCIs. As shown in Fig. 6(a) and (b), TRFs and spatial patterns of both paradigms exhibit remarkable similarity. In this case, the TRFs of the WN paradigm are bandpass-filtered into 8-15.8 Hz to ensure that both paradigms share a common frequency band. Therefore, the linear modeling method can also be applied to construct SSVEP by the convolution of the TRFs and the sinusoidal stimulus sequences.

We further showed that the constructed SSVEP templates are close to the real SSVEP response and thus can be used to boost identification performance, surpassing FBCCA. The comparative results in Fig. 6(c) confirms the efficacy of SSVEP reconstruction, as both amplitudes and phases are consistent between the real and the reconstructed signals. The target identification results further demonstrate that the linear modeling method outperforms FBCCA. It can be inferred from Fig. 6(d) that the highest average ITR is 157.70±35.49 bpm at 1.00 s with the linear modeling method, significantly higher than that with FBCCA, which is 134.38±34.95 bpm at 1.25 s (P<0.001 by two-way ANOVA). To sum up, the application of a single-target calibration phase can lead to more precise spatial-temporal pattern construction, thereby further improving the target identification performance of SSVEP-BCI.

## Conclusion

To address the issue of the need for extensive calibration and the low performance of cVEP-BCI, we introduce a single-target phase to significantly reduce the calibration time to a minute. Leveraging the novel WN cVEP-BCI paradigm, we propose the linear modeling and transfer learning methods for WN cVEP-BCI under minimal calibration, achieving an impressive ITR of 250.16 bpm. Moreover, we demonstrate that cVEP-BCI can achieve performance comparable to SSVEP-BCIs with minimal calibration. Our methods also show substantial progress in handling an increased number of targets, developing asynchronous systems, and enhancing the performance of SSVEP-BCI under limited calibration. We believe that our contributions will play a pivotal role in advancing the sustainability and effectiveness of cVEP paradigms.